# Predictive Analysis of CFPB Consumer Complaints Using Machine Learning

Dhwani Vaishnav, Manimozhi Neethinayagam, Akanksha Khaire, Jongwook Woo

Department of Information Systems, California State University Los Angeles
dvaishn2@calstatela.edu, mneethi@calstatela.edu, akhaire3@calstatela.edu, jwoo5@calstatela.edu

**Abstract**: This paper introduces the Consumer Feedback Insight & Prediction Platform, a system leveraging machine learning to analyze the extensive Consumer Financial Protection Bureau (CFPB) Complaint Database, a publicly available resource exceeding 4.9 GB in size. This rich dataset offers valuable insights into consumer experiences with financial products and services. The platform itself utilizes machine learning models to predict two key aspects of complaint resolution: the timeliness of company responses and the nature of those responses (e.g., closed, closed with relief etc.). Furthermore, the platform employs Latent Dirichlet Allocation (LDA) to delve deeper, uncovering common themes within complaints and revealing underlying trends and consumer issues. This comprehensive approach empowers both consumers and regulators. Consumers gain valuable insights into potential response wait times, while regulators can utilize the platform's findings to identify areas where companies may require further scrutiny regarding their complaint resolution practices.

## 1. Introduction

The Consumer Financial Protection Bureau (CFPB) is a U.S. government agency responsible for ensuring banks and other financial institutions treat consumers fairly. The CFPB maintains a publicly available Consumer Complaint Database [1]. This ever-growing CFPB Complaint Database offers a rich resource for understanding consumer experiences in the financial marketplace. This paper introduces a Consumer Feedback Insight & Prediction Platform that leverages machine learning models trained on consumer complaints data from 2007 to April 2024, the platform extracts actionable insights, predicting company response times and the nature of complaint resolution. Furthermore, it utilizes topic modeling to identify recurring themes within complaints, revealing prevalent consumer issues. This information equips both consumers and regulators with valuable tools. Consumers gain insights to make informed decisions, while regulators can leverage the platform to prioritize their efforts, ultimately fostering a fairer financial marketplace.

## 2. Related Work

The field of customer complaint analysis using machine learning is rapidly evolving, offering significant potential for improved customer service. Our work builds upon this foundation, drawing inspiration from several key studies.

Singh et al. (2023) explored the application of machine learning, specifically Logistic Regression and Support Vector Machines (SVM), for analyzing and predicting CFPB customer complaint data [2]. Their research demonstrates the effectiveness of this approach, with SVM achieving slightly better performance.

Li et al. (2023) focused on applying machine learning models specifically to predict complaint outcomes at Wells Fargo [3]. Their findings suggest that a Random Forest model can achieve high accuracy in predicting different complaint resolution paths.

While not directly related to machine learning, the CFPB's Consumer Response Annual Report 2023 underscores the importance of analyzing consumer complaint data for regulatory purposes [4]. The report details how the CFPB utilizes various techniques like text analytics and data visualization to monitor risks, assess company performance, and identify trends within the financial sector.

These studies all highlight the growing adoption of machine learning for customer complaint analysis. Our work expands upon this existing research in two keyways.

First, we aim to predict not only the likelihood of a timely response but also the nature of the company response itself. This broader prediction scope provides a more comprehensive picture of the complaint resolution process and empowers both consumers and regulators with additional insights.

Second, we incorporate topic modeling using Latent Dirichlet Allocation (LDA) to identify recurring themes within complaints. This offers valuable insights into broader industry trends and areas of concern, allowing for targeted interventions and improvements across the financial sector.

## 3. Specifications

The Consumer Complaint Database is a collection of complaints about consumer financial products and services that the CFPB receives from consumers. The dataset contains detailed information about each complaint, including the date of submission, the consumer's zip code, the type of financial product or service being complained about, and the nature of the complaint. The dataset is continuously updated and as of the date we downloaded, it's of the size 4.9 GB. It contains complaints data from 2011 to April 2024.

Below Table 1 shows files and size of the files from dataset.

*Table 1 Data Specification*

| Data Set Size | 4.9 GB |
|---|---|
| Number for files | 1 |
| Content Format | JSON |

The Table 2 below shows the specification for Oracle cluster we are using and pyspark specification for our paper.

*Table 2 H/W Specification*

| Number of nodes | 5 (2 master nodes, 3 worker nodes) |
|---|---|
| CPU speed | 1995.312 MHz |
| Storage | 800 GB |

## 4. Workflow

To develop the Prediction Platform, we adopted a structured workflow as shown in the *Figure 1* encompassing data preparation, model training, and evaluation. Initially, we acquired the CFPB Complaint Database and performed meticulous data preprocessing. This involved removing irrelevant columns, handling missing values, and employing frequency encoding for categorical features like company, issue, and state. This reduced data complexity and optimized it for machine learning algorithms. Subsequently, the preprocessed data was divided into training (70%) and test (30%) sets. The training set provided the foundation for model training. We utilized appropriate machine learning algorithms to predict two key aspects of complaint resolution: timely company response and the nature of the company's complaint resolution. Feature importance analysis was then conducted to understand the most influential factors for each prediction. To assess model performance, we employed metrics tailored to the specific prediction tasks. For the binary classification task of predicting timely company response, we utilized Area Under the Curve (AUC) in addition to the precision and recall. For the multiclassification task of predicting the nature of the company response (e.g., closed, closed with explanation, etc.), we evaluated the models using precision and recall assessing their ability to accurately classify different outcomes. This comprehensive evaluation ensured the robustness of the Consumer Feedback Insight & Prediction Platform.

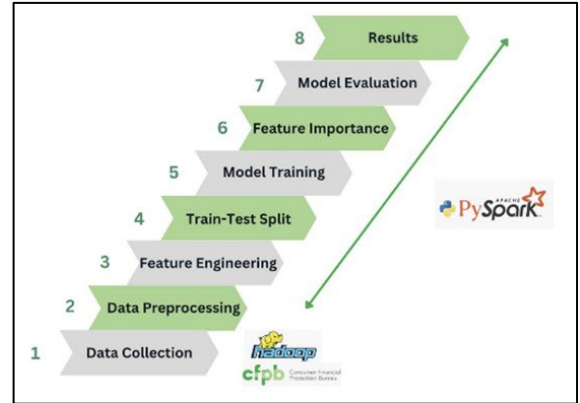

*Figure 1 Workflow*

## 5. Data Challenges and Preprocessing Techniques

Our raw complaint data presented two key challenges: imbalanced target variables and a high-cardinality features. Target variables for both timely response prediction and company response prediction exhibited significant class imbalances. Oversampling and, for company response, under sampling techniques were employed to create more balanced training sets. Additionally, the "Company" feature with its 7,000 unique values required attention. Frequency encoding tackled this challenge by transforming company names into numerical values based on their frequency within the dataset, effectively reducing complexity and improving model performance. These preprocessing steps ensured robust modeling and optimal results for analyzing and predicting consumer complaint outcomes.

## 6. Machine Learning

This section explores the use of machine learning to predict two key consumer complaint outcomes: timely responses and the nature of the company response.

### 6.1 Predicting Timely Responses: Binary Classification

The first aspect of our platform focuses on predicting whether a company will respond to a complaint within a designated timeframe. This information can be valuable for both consumers, who gain insights into potential response wait times, and regulators seeking to identify areas where companies may be exhibiting slow response patterns.

The model was trained on a set of **features** including Company Name, Product Category, Complaint Issue, State of Complaint Origin and Date Sent to Company.

The **target variable** is a binary indicator denoting "timely response" (Yes/No).

To achieve this prediction, we evaluated the performance of three machine learning algorithms: Gradient Boosted Trees (GBT), Support Vector Machines (SVM), and Logistic Regression (LR).

*Table 3 Comparison of Models for Binary Classification*

| Metric | GBT | LR | SVM |
|---|---|---|---|
| Precision | 0.85 | 0.69 | 0.69 |
| Recall | 0.93 | 0.96 | 0.57 |
| AUC | 0.94 | 0.87 | 0.87 |
| Computing Time | 35.45 Min | 31:35 Min | 47:21 Min |

As shown in *Table 3*, Gradient Boosted Trees (GBT) achieved the best overall performance across all metrics. While Logistic Regression demonstrated a high recall for "Yes" responses (indicating a good ability to identify timely responses), it missed many "No" responses. Support Vector Machine (SVM) exhibited comparable performance to Logistic Regression. Gradient Boosted Trees offered a superior balance, achieving very good results for "Yes" responses with a slight trade-off on "No" responses. The computing time for GBT training was higher than LR or SVM, but still within reasonable limits. These findings suggest that Gradient Boosted Trees is the most effective model for predicting timely responses to consumer complaints based on the features used.

## 6.2 Predicting Compony Responses: Multiclass Classification

This section explores predicting the nature of a company's response to a consumer complaint using a multiclass classification approach. This information can be valuable for understanding potential complaint resolution pathways and informing customer service strategies. These categories can include outcomes such as "closed with explanation," "closed with monetary relief," or "closed with relief." This classification task is considered multiclass as there are more than two possible response categories.

The model was trained on a set of **features** including Company Name, Product Category and Complaint Issue.

The **target variable** for this task is "company_response" with 8 unique categories:

- Closed with explanation
- Closed with non-monetary relief
- In progress
- Closed with monetary relief
- Closed without relief
- Closed
- Untimely response
- Closed with relief

To achieve this prediction, we evaluated the performance of two machine learning algorithms: Random Forest (RF) and Decision Tree (DT).

*Table 4 Comparison of Models for Multiclass Classification*

| Class | RF | | DT | |
|---|---|---|---|---|
| | Precision | Recall | Precision | Recall |
| Closed with explanation | 0.73 | 0.89 | 0.8 | 0.83 |
| Closed with non-monetary relief | 0.65 | 0.29 | 0.63 | 0.3 |
| In progress | 0.66 | 0.49 | 0.66 | 0.5 |
| Closed with monetary relief | 0.75 | 0.54 | 0.73 | 0.57 |
| Closed without relief | 0.41 | 0.49 | 0.45 | 0.66 |
| Closed | 0.42 | 0.75 | 0.51 | 0.55 |
| Untimely response | 0.55 | 0.36 | 0.51 | 0.56 |
| Closed with relief | 0.73 | 0.93 | 0.74 | 0.95 |
| Computing Time | 26.52 min | | 29. 33 min | |

As shown in *Table 4*, evaluation of both Random Forest (RF) and Decision Tree (DT) models revealed their effectiveness in classifying different categories of company responses. Notably, both models exhibited high recall scores (93% for RF and 95% for DT) for identifying instances of "closed with relief," indicating a robust capability for recognizing this outcome. For "closed with monetary relief," both RF (54%) and DT (57%) demonstrated a discernible capacity for identification, albeit with lower recall scores compared to "closed with relief."

An interesting finding is the trade-off observed in classifying "closed with explanation." While RF achieved a high recall score (88%), its precision (83%) suggests a higher rate of false positives compared to DT (83% recall, 90% precision). This implies that DT might miss some instances of "closed with explanation," but it produces more accurate classifications overall for this category.

Both models displayed minimal computational time, ensuring efficient processing. However, considering the slight advantage in recall scores and precision for key categories like "closed with explanation," the Decision Tree model emerged as the slightly more suitable choice for this multiclass classification task.

*Figure 2* shows the confusion matrix for Decision Tree Algorithm evaluation results. This confusion matrix offers valuable insights into our model's performance, especially its ability to distinguish between true classes and its predictions. This is evident from the high values along the diagonal of the matrix, indicating a clear separation between predicted and actual classes. This suggests that our model is effectively classifying data points.

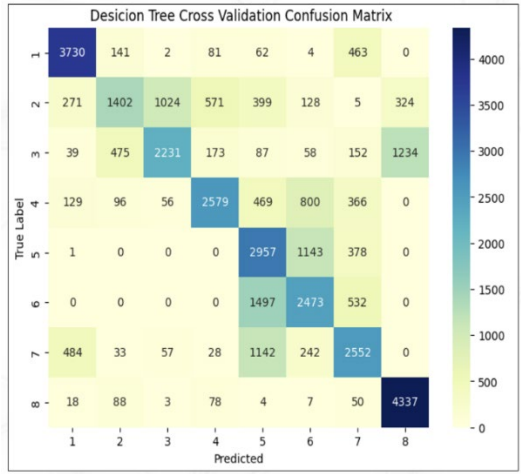

*Figure 2 Decision Tree Confusion Matrix*

### 6.3 Leveraging LDA for Topic Discovery in Consumer Complaints

Beyond classification models, we leverage Latent Dirichlet Allocation (LDA) to uncover underlying thematic structures within consumer complaint narratives. This approach, known as topic modeling, offers valuable insights into prevalent consumer financial concerns and areas requiring potential regulatory focus. LDA acts as a machine learning technique that identifies latent topics within a vast collection of documents, in this case, the 1.7 million consumer complaint narratives. By analyzing the most frequent words and phrases associated with each topic, we can gain a deeper understanding of the challenges consumers face.

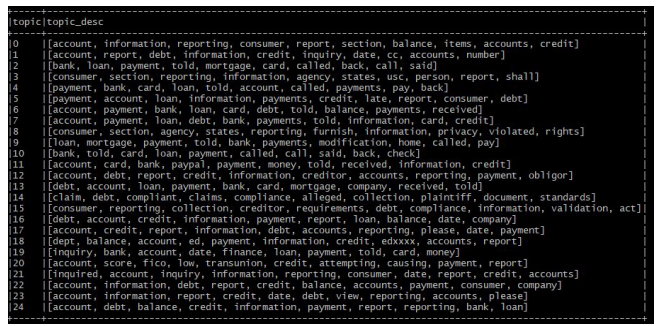

*Figure 3 LDA Topic Discussions*

For instance, as shown in *Figure 3*, LDA might reveal a prominent topic related to "Consumer Credit Reporting" (Topic 0), encompassing issues like credit report inaccuracies or disputes regarding credit inquiries. Similarly, topics like "Banking & Loans" (Topic 2) or "Mortgage-Related Matters" (Topic 9) could point towards challenges with loan applications, servicing, or potential unfair lending practices. These identified themes can inform targeted investigations by regulators and consumer advocacy groups, while also serving as a foundation for prioritizing CFPB regulations based on the most frequent complaint topics. By continually monitoring and analyzing these topics over time, we gain insights into evolving consumer financial concerns and tailor interventions accordingly.

### 7. Conclusion

This study explored machine learning's potential for analyzing CFPB consumer complaint data. Leveraging a public dataset and Apache Hadoop/PySpark for processing, we investigated key complaint outcomes: timely company responses and the nature of company's response. Gradient Boosted Trees excelled at predicting timely responses, while Decision Tree outperformed in classifying response types, particularly frequent categories. Additionally, Latent Dirichlet Allocation (LDA) uncovered thematic structures within complaints, revealing common issues like credit reporting, banking, and mortgages. These findings highlight the potential of machine learning to empower regulators, policymakers, and financial institutions.